\documentclass[smallcondensed]{svjour3} 

\usepackage[utf8]{inputenc}
\usepackage{booktabs}
\usepackage{amsmath}
\usepackage{amssymb}
\usepackage{natbib}
\usepackage{algorithm}
\usepackage{algorithmic}

\usepackage{multirow} 
\usepackage{subcaption} 
\usepackage{dsfont} 
\usepackage{rotating}
\usepackage{lscape}
\usepackage{todonotes}
\usepackage{threeparttable}
\newcounter{constraint}
\usepackage{colortbl}

\newcommand{\nextconstraint}{\refstepcounter{constraint}\arabic{constraint}.}
\DeclareMathOperator*{\argmin}{argmin}
\usepackage{xparse} 
\NewDocumentCommand{\rot}{O{45} O{1em} m}{\makebox[#2][l]{\rotatebox{#1}{#3}}}%

\AtBeginDocument{%
  \providecommand\BibTeX{{%
    \normalfont B\kern-0.5em{\scshape i\kern-0.25em b}\kern-0.8em\TeX}}}

\begin{document}

\title{Capacitated spatial clustering with multiple constraints and attributes}
\subtitle{A case study in edge server placement}

\author{Leena Ruha* \and 
Tero Lähderanta* \and 
Lauri Lovén*\thanks{*These authors contributed equally.} \and 
Teemu Leppänen \and
Jukka Riekki \and 
Mikko J. Sillanpää
}
 
\authorrunning{Ruha, Lähderanta, Lovén et al.}

\institute{L. Ruha \at
           Natural Resources Institute Finland, Oulu, Finland \\
           \email{leena.ruha@luke.fi}
           \and
           T. Lähderanta \at
           University of Oulu, Oulu, Finland \\
           \email{tero.lahderanta@oulu.fi}
           \and
           L. Lovén \at 
           University of Oulu, Oulu, Finland \\
           \email{lauri.loven@oulu.fi}
           \and
           T. Leppänen,  J. Riekki, M. J. Sillanpää \at
           University of Oulu, Oulu, Finland
}

\date{Submission date: 1.3.2021}
%
%

\maketitle
\begin{abstract}

 
Capacitated spatial clustering, a type of unsupervised machine learning method, is often used to tackle problems in compressing, classifying, logistic optimization and infrastructure optimization. Depending on the application at hand, a wide set of extensions may be necessary in clustering. 
In this article we propose a number of novel extensions to PACK that is a novel capacitated spatial clustering method. These extensions are relocation and location preference of cluster centers, outliers, and non-spatial attributes.
The strength of PACK is that it can consider all of these extensions jointly. We demonstrate the usefulness  PACK with a real world example in edge computing server placement for a city region with various different set ups, where we take into consideration outliers, center placement, and non-spatial attributes. Different setups are evaluated with summary statistics on spatial proximity and attribute similarity. As a result, the similarity of the clusters was improved at best by 53\%, while simultaneously the proximity degraded only 18\%. In alternate scenarios, both proximity and similarity were improved. The different extensions proved to provide a valuable way to include  non-spatial information into the cluster analysis, and attain better overall proximity and similarity. Furthermore, we provide easy-to-use software tools (rpack) for conducting clustering analyses.


  

\end{abstract}

%


\section{Introduction}

Clustering,  one of the most important tasks and techniques in data mining and unsupervised machine learning, refers to the unsupervised classification of patterns into groups. Its primary goals include preprocessing, compressing, classifying and gaining insight into the data \citep{Celebi14, Patel16, Grubesic14}.

This work focuses on partitional (also referred to as non-hierarchical) clustering. Partitional clustering aims to partition the data set into non-overlapping subsets such that each data point is in exactly one subset and each cluster can be represented by one point, referred to as a prototype \citep{Jin10,Xiao12}. Partitional clustering is especially useful in applications which use the partitions for further analyses. As an example, direct marketing campaigns often start with segmenting customers into groups \citep[see further examples in][]{Banerjee06}.

Further, clustering often involves a spatial dimension with geographic information related to the studied phenomenon. Such a setting, referred to as spatial clustering, requires an appropriate and meaningful treatment of space, spatial relationships, and the attributes of locations \citep{Grubesic14}. We focus in particular on partitional spatial clustering, where points of interest are partitioned into disjoint clusters. Partitional spatial clustering is used, for example, for the spatial analysis of Internet of Things (IoT) sensor data~\citep{lee2015}. 
Partitioning the sensor data into local clusters can help in finding local features of the observed phenomena, and in distributing the computational burden of the data analytics, especially if local or edge-based computing capacity is available \citep{yi2015,xu2017,Loven2019c}.

In this article, we present a novel capacitated partitioning approach that jointly considers constraints related to the locations of the cluster centers, outliers, and non-spatial attributes. Constraints for center placement corresponds to a problem where the location of some cluster heads are known. In more detail, instead of knowing the location of some centers exactly, there may be more or less vague prior information about their location. On the other hand, if some center locations are fixed, it may be possible relocate such centers under certain circumstances. Contrary to related works, we consider such constraints jointly with capacitated clustering.

Outlier points, on the other hand. may distort the clustering structure. Thus it may be beneficial to allow some points to be assigned as outliers, i.e. not assigned to any cluster head or facility. However, we know of now work which considers outliers jointly with capacitated clustering.

Finally, the spatial distribution of the points may not be the only determinative factor in the clustering problem. One example is found in customer segmentation where customers are divided into clusters not only based on their spatial location, but also the total number of purchases and the geodemographic lifestyle class \citep{Brimicombe2007}. Again, to the best of our knowledge, no related work considers such non-spatial attributes in conjunction with capacitated spatial clustering, or the other extensions we presented above.

We demonstrate the feasibility and importance of the joint consideration of the above extensions with an experiment in edge computing. Edge computing is recognized as a paradigm shift from centralized, cloud-based processing towards distributed computing near the sources of the data~\citep{satyanarayanan2017emergence}. A key part of, e.g., the the upcoming 5G and beyond mobile networks, edge promises more bandwidth, low latency and privacy improvements through virtualization of cloud-based applications into the network infrastructure, close to the end user devices. However, distributing application components across the communication networks introduces challenges in the orchestration and provisioning of resources across the opportunistic environment in a massive-scale. Machine learning is proving to provide feasible solutions to these challenges~\citep{Park2018, Loven2019edgeai1, Xu2020, Zhu2020a}. 

Our experiment deals with edge server placement~\citep{Lahderanta19, Loven2020a}. While topical in its direct connection to edge computing, edge server  placement also provides a suitable experimental setup for evaluating the multi-objective nature of our proposed method. Indeed, edge server placement requires simultaneously fulfilling a number of requirements. First, edge servers need to be placed in close proximity to the users and access points, minimizing latencies. Second, allocation of user workload should consider the capacities of the edge servers, ideally ensuring a balanced workload division between the servers. Third, placement may require the consideration of preferred locations (e.g. airports or shopping centres), outliers (e.g. access points whose workload can be readily uploaded to cloud), and non-spatial attributes (e.g. application usage profiles). We use the Shanghai Telecom data set \citep{Guo19, Wang19, Xu19} in our evaluations. The data set contains over 7.2 million sessions from mobile phones to 2732 base stations in the city of Shanghai in a six-month period. 


We also provide an extensive, open-source toolkit for conducting the proposed analyses\footnote{https://github.com/terolahderanta/rpack}. We refer to the proposed approach as PACK (PlAcement with Capacitated K-family\footnote{Here, k-family refers to the family of different variants of K-mean, i.e. K-mean, K-medoid, and K-median}). We have proposed earlier versions of PACK, with fewer features, in our previous articles for placing edge servers in the city of Oulu \citep{Lahderanta19,Loven2020a} and maternity hospitals in Finland \citep{Huotari20}. However, compared with the earlier versions, we make some important enhancements. In short, our contributions are as follows:
\begin{enumerate}
    \item We consider novel extensions for center placement in spatial clustering, namely, relocating fixed centers, and a preference for certain locations over the others. 
    \item We consider novel extensions for capacitated spatial clustering, namely, outliers, i.e., isolated points in the data that may distort the clustering structure, and non-spatial attributes of the points of interest.
    \item We examine the effects of the above extensions and their combinations in edge computing server placement, with a real-world data set of a mobile network in the city of Shanghai, China.
\end{enumerate}

The rest of the article is organized as follows. In Section \ref{related_work} we take a detailed look on the existing literature on capacitated clustering problems. In Section \ref{method} we present the PACK method and its extensions. In Section \ref{evaluation} we demonstrate PACK with an experiment on edge server placement in multiple setups. In Section \ref{discussion} we highlight the main findings of our evaluations.


\section{Related Work}
\label{related_work}



We look at related work along the axes of center placement, balance, outliers, and non-spatial attributes.

\textbf{Center placement.} 
Sometimes the locations of the cluster centers need to be constrained, often to coincide with the data points. Such location-constrained algorithms are referred to as \textit{actual point-prototype-based clustering algorithms} \citep{Xiao12}. One of the most common of such algorithms is the \textit{partitioning around medoids} (PAM) algorithm \citep{Kaufman1990}. 
However, we found neither a study observing that fixed centers may result in a sub-optimal partition, nor offering ways to relocate some of the fixed centers to produce better results.

Finally, \cite{Lahderanta19} assigned some demand locations with a location preference, indicating they should have a center close to them. However, they did not consider that location preference can also be considered for cluster centers.


\textbf{Capacity constraints.} Clustering is often used for dissecting data into separate groupings for further processing or analysis.  Such cases often control the size of clusters, such that too large or too small clusters are undesirable. These types of clustering methods can be partitioned into two groups, balance constrained and balance driven approaches \citep{Malinen14}. Here we investigate the related work on balance constrained clustering.

In balance constrained clustering, the cluster size balance must be met, and the minimization of distances is only a secondary criterion. This is typically achieved with an additional constraint to the capacity of a cluster. Several approaches have been proposed. Balance constrained include \citet{Banerjee06},  \citet{Malinen14} and \citet{Elliott11}. Other constrained approaches include those by \citet{Hu18}, \citet{Zhu10} and \citet{Ganganath14}, where \citet{Hu18} constrain the size of the clusters only from below, and \citet{Zhu10} and \citet{Ganganath14} allow predefined size constraints to vary between clusters. \citet{LI18} and \citet{Liu17} studied balance-driven clustering with a soft constraint that penalizes large deviations in cluster sizes, \citet{Gupta17} recently provided survey on balanced data clustering.

Hard capacity constraints have also been proposed in the clustering framework by \citet{Mulvey84}, who assumed that the weights $w_i=1$ and coefficients $a_i=1$ for all $i$ and referred to the problem as the \textit{capacitated clustering problem} (CCP). Later, \citet{Negreiros06} used squared Euclidean distances in a continuous setting and referred to the problem as \textit{capacitated centered clustering problem} (CCCP). Another possibility for combining balance and capacity constraints is to apply both a lower and an upper limit for the used capacity \citep{Borgwardt17}. However, no studies consider capacitated spatial clustering in conjunction with the extensions in the Section \ref{method}.

\textbf{Outliers.} If a point does not seem to belong to any cluster, a forced assignment may distort information and the interpretation of the cluster it is assigned to \citep{Tseng07}. It may thus be necessary to allow a point not to be a member of any center. In partitional clustering, such points that are left unassigned are referred to as outliers. Further, a partition where every point is assigned is referred to as complete clustering, and a partition where some points may be left unassigned is referred to as partial clustering \citep{Steinbach19}. 


 
Outliers could be removed prior to the clustering by using some outlier detection method. However, the clustering structure may be highly dependent on the outliers, and on the other hand, the determination of the outliers depends on the clustering structure. Identifying the outliers should thus not be a separate step, but a part of the clustering process \citep{Liu19b}.

Outliers can be identified by determining the number of outliers prior to analysis, and iteratively assigning the most remote points as outliers. Such an approach has been proposed by, e.g., \citep{Chawla13}. 
\citet{Whang15} combine outlier analysis with overlapping clustering by proposing an overlapping K-means algorithm where a fixed number of points is assigned to at least one cluster and a fixed number is not assigned at all.
Alternatively, outliers can be identified by keeping track of the distances of points to their nearest center.  \citet{Olukanmi17} determine such threshold iteratively.


A third alternative is to formulate the problem as a bi-criteria optimization, where a fixed penalty  is paid \citep{Tseng07} for each outlier point. Thus, the outliers can be considered to be assigned to a $k+1$'th cluster, with a penalty added to the loss function.
\citet{Tseng07} further give this bi-criterion approach a model-based interpretation, pointing out that the penalty term  corresponds to assuming that the noise points are uniformly distributed, i.e., that they emerge from a homogeneous Poisson process.

None of the related work, however, considers outliers in conjunction with spatial capacitated clustering.

\textbf{Non-spatial attributes.} In many real world scenarios it is necessary to consider also non-spatial attributes. Such an approach is referred to as \textit{dual clustering} \citep{Lin2005}. 

Dual clustering methods can be divided into two subcategories: the hybrid distance measure approach, and the separate clustering operations approach \citep{Zhu2020}. In the first approach, the spatial and non-spatial attributes are included with a distance measure, which consists of both a spatial and a non-spatial similarity measure, each weighted depending on the problem in question. \citet{Lin2005} further develop the Interleaved Clustering-Classification algorithm with a hybrid-distance measure, with weight $w$ in the spatial attributes and weight $1 - w$ in the non-spatial attributes. \cite{Zhang2007} aim to obtain non-overlapping clusters by embedding a penalized spatial distance measure (PSD). They compare PSD to tje standard 2-dimensional Euclidean distance with only spatial attributes and to 3-dimension Euclidean with spatial and non-spatial attributes.

Again, however, we found no related work that considers non-spatial attributes jointly with capacitated clustering.

\textbf{PACK.} We have proposed earlier versions of our approach, named PACK (see Section~\ref{method}), with fewer extensions, in \citet{Lahderanta19} and \citet{Loven2020a}. Newest version of PACK includes many of the approaches presented in the related work and most importantly PACK allows the use of multiple extensions simultaneously, which  none of the previous studies have implemented. For highlighting the novelty of this article, we summarize the extensions introduced in the different versions in Table~\ref{Comparison}.
\begin{table}[ht]
  \caption{\textbf{The extensions included in different versions of the PACK algorithm.} Extensions newly introduced in each version are highlighted with green color.}
    \begin{center}
    \begin{tabular}{l l l l l }
    \toprule
        \textbf{Feature}  &  &  \textbf{PACK 1}  & \textbf{PACK 2} & \textbf{PACK 3} \\
    \midrule
        Center  & Fixed centers &  - & \textcolor{green}{\checkmark} & \checkmark\\
                & Released centers & - & -  & \textcolor{green}{\checkmark}\\
    \midrule
        Location preference & Point-based & \textcolor{green}{\checkmark} & \checkmark & \checkmark\\
        & Center-based & - & - & \textcolor{green}{\checkmark}\\
    \midrule
    Non-spatial attributes & & - & - & \textcolor{green}{\checkmark}\\
    \midrule
        Capacity limits &  & \textcolor{green}{\checkmark} & \checkmark & \checkmark\\
    \midrule
        Membership  & Hard & \textcolor{green}{\checkmark} & \checkmark & \checkmark \\
                    & Fractional & \textcolor{green}{\checkmark} & \checkmark & \checkmark\\
                    & Overlapping
                    & \textcolor{green}{\checkmark}  & \checkmark & \checkmark\\ 
    \midrule
        Outliers & & - & - & \textcolor{green}{\checkmark}\\  
    \bottomrule
    \multicolumn{4}{l}{\footnotesize{PACK 1: \citet{Lahderanta19}}}\\
    \multicolumn{4}{l}{\footnotesize{PACK 2: \citet{Loven2020a}}}\\
    \multicolumn{4}{l}{\footnotesize{PACK 3: This paper}}  
\end{tabular}
\end{center}
\label{Comparison}
\end{table}

\section{The PACK method}\label{method}

We propose a capacitated spatial clustering method which introduces a number of novel extensions, while covering a wide range of extensions earlier considered only separately. We refer to the proposed approach as PACK (PlAcement with Capacitated K-family). 

\begin{center}
\begin{table}
\centering
\caption{The symbols utilized in PACK.}
\label{symbols}
\begin{tabular}{l l l l}
\toprule
&\textbf{Symbol} &\textbf{Range} &   \textbf{Description}  \\
\midrule
\multirow{2}{*}{\rot[90]{\textbf{Inputs}}}  
&$x_i, \; i=1,\dots,n$ & $\mathds{R}^2$ & the spatial coordinates of point $i$\\
&$\theta_i, \; i=1,\dots,n$ & $\mathds{R}^Q$ & non-spatial attributes of point $i$\\
&$x^*_i = \{x_i, \theta_i\}, \; i=1,\dots,n$ & $\mathds{R}^{2 + Q}$ & the location and the non-spatial attributes of point $i$\\
&$Q$ & $\mathds{N}$ & dimension of non-spatial attributes\\
&$w_i$ & $[0, \infty[$ & weight of point $i$ \\
&$d(\cdot, \cdot)$ & $[0, \infty[$ & distance between two locations \\
&$\gamma_i$ & $[0, \infty[$ & location preference of point $i$ \\
&$w'_i=\gamma_i+w_i$  & $[0, \infty[$ & weight of point $i$, corrected by its location preference \\
&$p_h, \; h=1,\dots, s$  & $\mathds{R}^2$ & a possible location for a center \\
&$m$   & $\mathds{N}$ & the number of centers with fixed location \\
& $f_o,\; o=1,\dots,m$   & $\mathds{R}^2$ & location of a fixed center \\
&$\lambda_o $  & $[0, \infty[$ & the penalty parameter of outliers \\
&$\lambda_f $ & $[0, \infty[$ & the penalty parameter for releasing a fixed center \\
&$L$   & $[0, \infty[$ & the lower capacity limit  \\
&$U$  & $[L, \infty[$& the upper capacity limit \\
&$a_i$   & $[0, \infty[$ & the weight used in the capacity constraints \\
&  & &(typically $a_i=w_i$) \\
\midrule
\multirow{2}{*}{\rot[90]{\textbf{Outputs}}}
&$A$  & $\{\mathds{R}^2\}$ &set of released centers  \\
&$t=|A|$  & $\mathds{N}$ & number of released centers \\
&$c_j, \; j=1,\dots,k$  & $\mathds{R}^2$& the spatial coordinates of center $j$ \\
&$c^*_j = \{c_j, \theta_i\}, \; i=1,\dots,n$ & $\mathds{R}^{2 + Q}$ & the location and the non-spatial attributes of center $j$\\
&$y_{i j}$ & $[0, 1]$ & membership of point $i$ to center $j$ \\
  \bottomrule
\end{tabular}
\end{table}
\end{center}
\subsection{Problem formulation}
\label{New_formulation}
Table~\ref{symbols} lists the symbols used. PACK minimizes the following objective function
\begin{equation}
\label{lossOurs}
\argmin_{c^*_j, y_{i j},A} \; \;\underbrace{\sum_{i = 1}^{n} \sum_{j = 1}^{k} w'_i d(x^*_i, c^*_j) y_{i j}}_{\text{distance}} + \underbrace{\lambda_o \sum_{i = 1}^{n} w'_i y_{i,k+1}}_{\text{outliers}} 
+  \underbrace{\phantom{\sum_{i = 1}^{n}}\lambda_f t\phantom{\sum_{i = 1}^{n}}}_{\text{released centers } }
\end{equation}
with the following constraints:

\vspace{0.5cm}
\begin{center}
\begin{tabular}{ l c}
\nextconstraint\label{const_centers} Center locations & $c_j \in \{p_1, p_2,... ,p_s\}, \quad \forall j,$  \\[0.5cm]
\nextconstraint\label{const_fixed} Fixed centers & $c_l =f_l, \quad l\in   \{1,\dots,m\} \setminus A$,  \\[0.5cm]  \nextconstraint\label{const_membership} Membership & $y_{i j} \in [0, 1], \quad \forall i,j$  \\[0.5cm] 
\nextconstraint\label{const_membership2} Membership &  $\sum_{j = 1}^{k} y_{i j} = 1 \quad \forall i$, \\[0.5cm]
\nextconstraint\label{const_capacity} Capacity constraints & $L \leq \sum_{i = 1}^{n} a_i y_{i j} \leq U  \;  \forall j$.
\end{tabular}
\end{center}
\vspace{0.5cm}

\textbf{Weights and location preference.}
PACK allows weighted data points and enables the inclusion of preferred cluster center locations. These preferred locations are incorporated into the problem by adding a parameter $\gamma_i>0$ to the weight of the preferred points $w_i$ i.e. the total weight $w'_i=w_i+\gamma_i$.  The larger the parameter $\gamma_i$, the more strongly the point $i$ attracts a center \citep{Ackerman12}.


\textbf{Fixed and released centers.} Applying constraint \ref{const_fixed}, $0 \leq m\leq k$ centers can be assigned to have predetermined locations $f_1,\dots,f_m$. Further, we propose a relaxation where a fixed center at a preassigned location can be released and relocated if a penalty $\lambda_f$ is paid. 
Thus, if $t$ fixed centers are released, the cost is $\lambda_f t$. A center is released if the release  decreases  the value of the objective function more than $\lambda_f$. 



\textbf{Distance metrics.}
PACK is agnostic towards the distance metric used. The objective function (\ref{lossOurs}) admits an arbitrary distance measure $d$. For example standard Euclidean distance and squared Euclidean distance can be used. 

Distance function $d$ takes four arguments: $\{x_i, \theta_i\}$ and  $\{c_j, \theta_j\}$, which correspond to the spatial coordinates and the non-spatial attributes of point $i$ and center $j$. Furthermore, in a case of dual clustering scenario a hybrid distance measure can be applied in PACK. This is typically done by dividing the distance measure to a sum of two distances: $d(x^*_i, c^*_j) = d(\{x_i, \theta_i\}, \{c_j, \theta_j\}) = \lambda_d d_{1}(x_i, c_j) + (1-\lambda_d) d_{2}(\theta_i,\theta_j)$, where $\lambda_d$ is the weight given to spatial distance measure $d_{1}$ and $(1-\lambda_d)$ is the weight given to non-spatial distance measure $d_{2}$. 

\textbf{Capacity.}
Both lower and upper capacity constraints can be applied for controlling the amount of weight assigned to a center. Depending on the upper and lower limits, this constraint can be used for both balancing the data and constraining only either too large or too small clusters. 

Further, PACK allows the capacity constraint and the loss function to have different constants ($a_i$ and $w_i$, respectively).  
This allows, for example, applying a location preference for the objective function but not for the capacity limits.

\textbf{Outliers.}
PACK identifies outliers by applying the bi-criterion approach \citep{Tseng07, Charikar01}.  
Instead of assigning a constant outlier penalty $\lambda_o$ for each point, we use a penalty that is proportional to the weight of the point, $\lambda_o w'_i$. As a result, the capacity limits permitting, a point is assigned as an outlier if $d(x^*_i,c^*_j) > \lambda_o$. 
Hence, the selection depends  on the distance to the clusters and not on the weight of the point. 

\subsection{The block coordinate descent algorithm} 
\label{BlockCoord}

\begin{algorithm}[t]
\caption{PACK-algorithm}
\textbf{Input:} $x^*_i, w'_i, k, n, \lambda_o, \lambda_f, f_j, j=1,\dots,m$ \\
\textbf{Output:} locations of the centers $c_j'$ and allocations of points to centers $y_{i j}'$, $j=1,\dots, k+1$\\
\label{alt_alg}
\begin{algorithmic}[1]
\FOR{$i = 1$ to $n$}
\STATE Initialize $c^*_j$, $j = 1,2,...,k$ using K-means++
\WHILE{$c^*_j$ changes}
\STATE Allocation-step: minimize (\ref{lossOurs}) with respect to $y_{i j}$ \label{allocation_step}
\STATE Location-step: minimize (\ref{lossOurs}) with respect to $c^*_j$ \label{location_step}
\STATE $S = $ the value of the objective function
\ENDWHILE
\IF{$S < S_{min}$ or $i = 1$}
\STATE $S_{min} \leftarrow S$
\STATE $c_j' \leftarrow c^*_j$
\STATE $y_{i j}' \leftarrow y_{i j}$
\ENDIF
\ENDFOR
\RETURN $c_j', y_{i j}'$
\end{algorithmic}
\end{algorithm}

PACK employs a block coordinate descent algorithm \cite{Tseng01} for minimizing (\ref{lossOurs}) with a fixed number of centers (Algorithm \ref{alt_alg}). On each iteration, a block coordinate descent algorithm minimizes the objective function with respect to a block of variables while holding other blocks of variables fixed at the values obtained in the previous iteration step \citep[see e.g.][]{Wright15}.
The two main steps, i.e., the blocks, for the algorithm are the \textit{allocation-step} (line \ref{allocation_step}), where the points are assigned to the centers, and the \textit{location-step} (line \ref{location_step}), where the centers are relocated based on the points assigned to them. The steps are iterated until convergence is reached. 

Further, as block-coordinate descent finds the local minima close to the initial values, PACK uses a number of initial values to find the global minimum. The initial values are obtained using the K-means ++-algorithm \citep{Arthur07} that spreads the initial locations of centers improving both the speed and the accuracy of the K-means method. 


\textbf{Allocation step.}
The allocation step minimizes the objective function (\ref{lossOurs}) with respect to $y_{i j}$. Constraints \ref{const_membership}, \ref{const_membership2} and \ref{const_capacity} are applied, and the locations of the centers $c^*_j$ are assumed to be fixed. 
If no capacity constraints are applied, this step corresponds to assigning each point to the nearest facility measured in terms of the chosen distance metric. In addition a point is assigned as an outlier if its distance to the nearest center exceeds the limit $\lambda_o$. On the other hand, if capacity constraints are applied, this step is an NP-hard integer programming. 
 



\textbf{Location step.}
Location step minimizes the objective function (\ref{lossOurs}) with respect to $c^*_j$'s, while keeping the allocations $y_{i j}$ fixed. In other words, each center is separately relocated given the points assigned to it. This step is omitted for fixed centers. 

In continuous space with the squared Euclidean distance metric, $c_j$ is the weighted mean of the points allocated to cluster $j$. 
In the discrete setting, $c^*_j$ is the point which minimizes the sum of pairwise distances among points assigned to $c^*_j$. 

If releasing fixed centers is allowed, a center is released and relocated if the sum of distances to the assigned points is reduced more than $\lambda_f$. Similarly, a previously released fixed center can be returned to its original location if the increase in the sum of distances is lower than the release penalty $\lambda_f$.


\textbf{Implementation.}
The algorithm is implemented as an open source R package called \textit{rpack} \citep{rpack} available on GitHub. The allocation step is run on Gurobi \citep{gurobi}, a fast optimizer package with R bindings freely available for academic use. If Gurobi is not available, PACK uses the lpSolve-package for R \citep{lpsolve} for the allocation step.  

\section{Evaluation with edge server placement}
\label{evaluation}

\begin{figure}
    \centering
    \begin{subfigure}[t]{0.65\textwidth}
    \includegraphics[width=\textwidth]{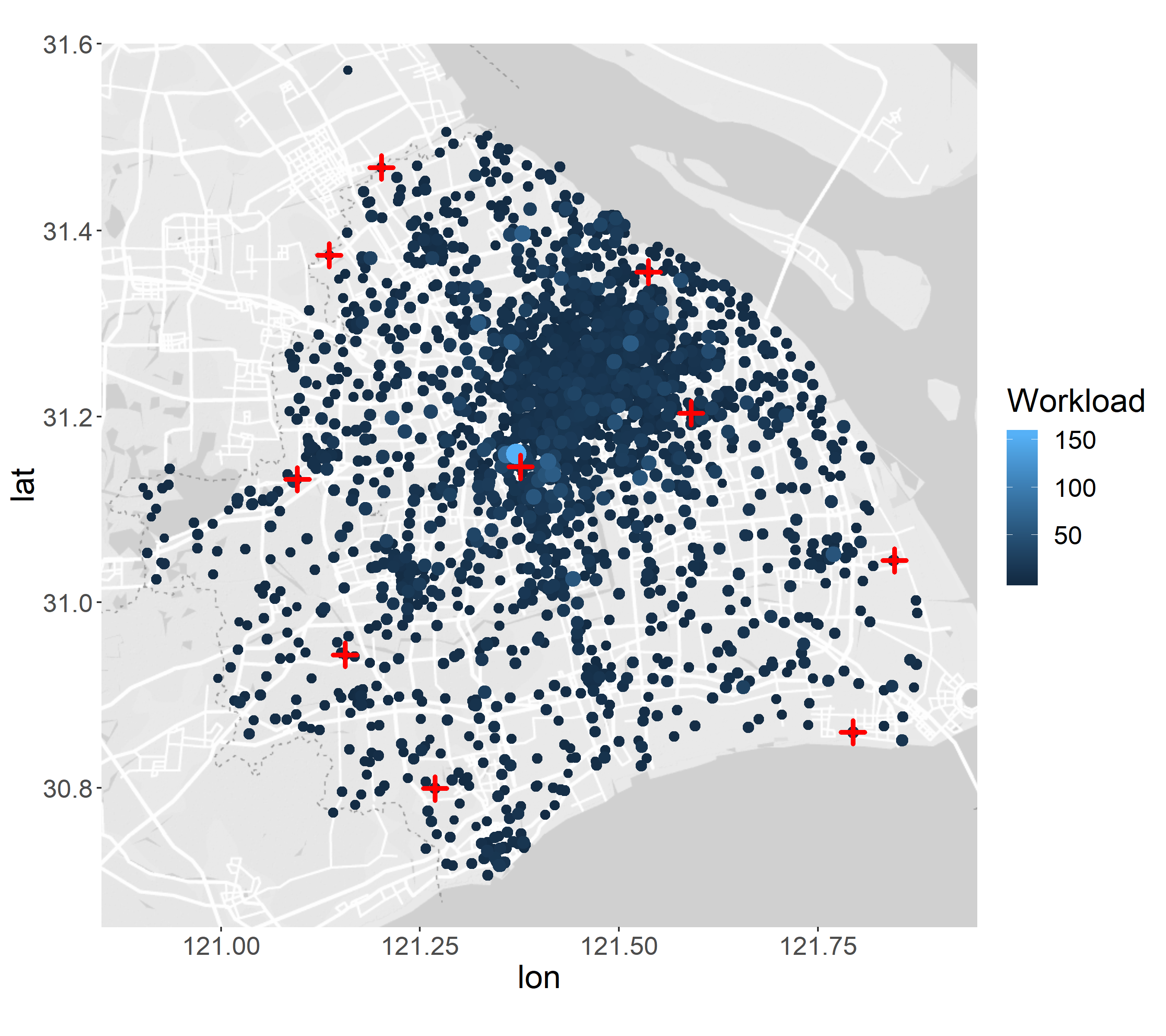}
    \end{subfigure}
    \begin{subfigure}[t]{0.65\textwidth}
    \includegraphics[width=\textwidth]{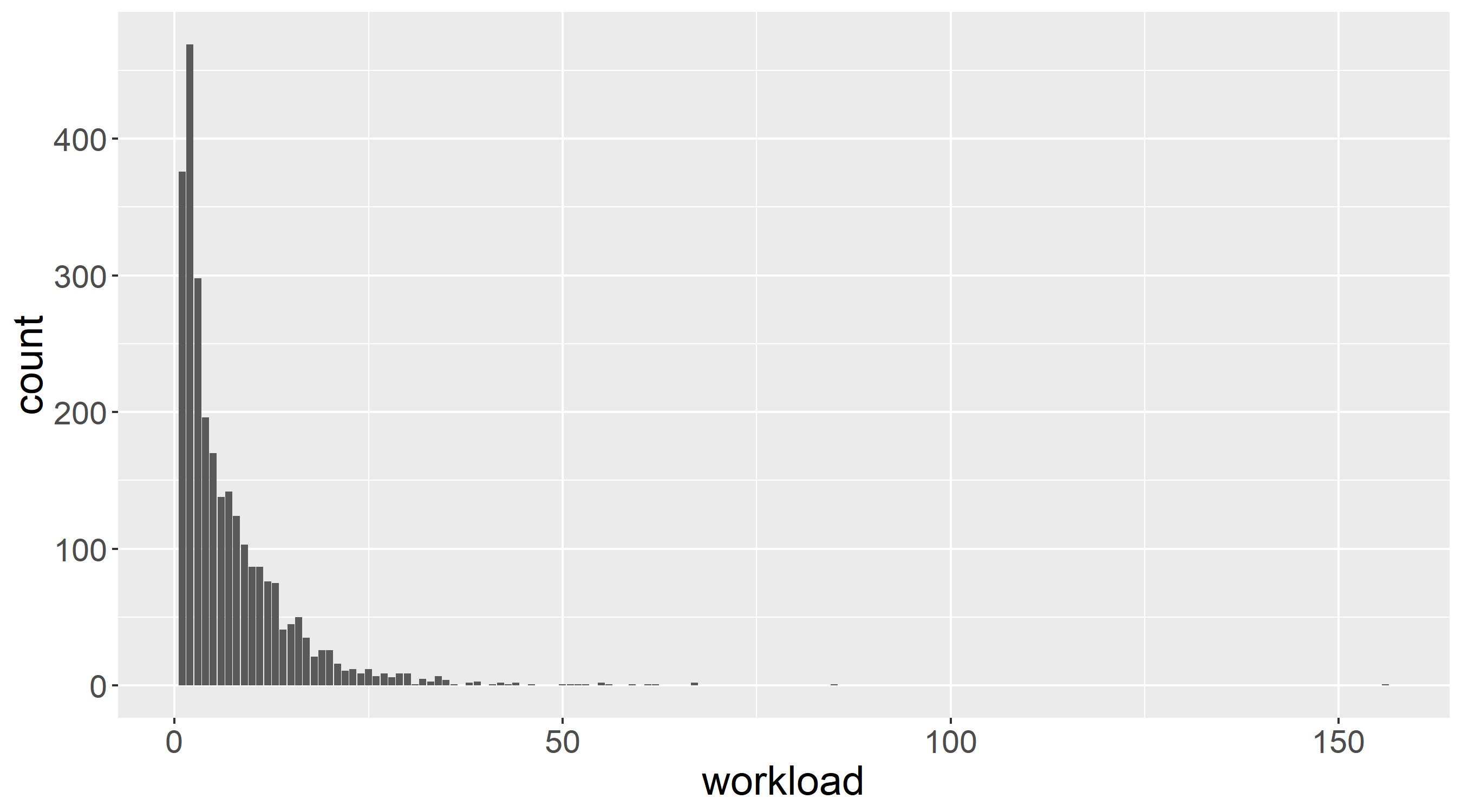}
    \end{subfigure}
    \caption{\textbf{Base stations in Shanghai area.} Left: the locations of base stations, color representing the workload and the red crosses the 10 fixed edge server locations. Right: the workload distribution of the base stations. }
    \label{fig_shanghai}
\end{figure}

We demonstrate the feasibility of PACK with an experiment focusing on the placement of edge servers in the city center of Shanghai, China \citep{Guo19, Wang19, Xu19}.  The Shanghai Telecom data set contains the locations of 2732 base stations in the Shanghai region and the mobile user connections to those base stations in a six-month period.  Spatial distribution of the base stations and their workloads are shown in Fig.~\ref{fig_shanghai}.  We demonstrate the use of capacity constrains, non-spatial attributes, fixed server location, releasing fixed servers, location preference and outliers.

The experiment optimizes the telecommunications infrastructure in Shanghai, placing \textit{edge computing servers} (ES) to reduce the communication latency experienced by the mobile phone users. Edge computing \citep{shi2016} refers to computing infrastructure that facilitates data processing directly at the points of interest. Deploying such infrastructure, characteristics such as radio coverage, the available telecommunications networks, and users' need for computational capacity, largely dictate the physical placement options of the servers \citep{Lahderanta19}. 


The goal of ES placement is to deploy a number of edge computing servers in the city region such that the latencies (i.e. distances) between the servers and the base stations assigned to each server are minimized. Each base station is assigned to exactly one ES, and the workload (i.e., weight)  of the base station, that is, its maximum number of concurrent users in the recorded data, is allocated to that server. 



Each ES, represented by a cluster center, has a limited computing capacity for the workload it can handle, represented by the sum of the weights of the base station locations in the corresponding cluster. Further, each edge server can only be placed at the same location as one of the base stations. Since the topology of the underlying network is unknown, we use geospatial distances to approximate the latencies between the base stations.

ES placement provides an ideal setting for studying the PACK extensions, as these extensions have clear real-world interpretations. We explore the effect of extensions in two scenarios:
\begin{enumerate}
    \item Adding the non-spatial attribute \textit{average session length} to the placement problem as an additional parameter and the inclusion of outliers. 
    \item Joint consideration of 1) fixed server locations, with ten servers already deployed or having preferred locations, and 2) including the average session length in the placement. 
\end{enumerate}

In our evaluations, we place $k = 38$ edge servers in the region. The number of servers was decided using the elbow method \citep{Yuan2019}. 

We use the average session duration as a proxy to end-user application usage profiles, which were unavailable in the open data set. \cite{Bohmer2011} list distinctly varying usage durations for different mobile applications, with e.g. an average of 68.11 seconds for news applications, and 114.25 seconds for game usage. Clustering similar application usage patterns for individual edge servers helps in, for example, caching and pre-loading of popular content to that ES, greatly reducing user latency and communication burden on the core network. 


Each base station has spatial coordinates in longitudes and latitudes $x_i$, and an average session length $\theta_i$, which we then combine into the vector $\{x_i, \theta_i\}$. We study the impact of average session duration by implementing a hybrid distance function $d$ which is a sum of the spatial and non-spatial component: $d(\{x_i, \theta_i\}, \{c_i, \theta_j\}) = \lambda_d d_{1}(x_i, c_j) + (1-\lambda_d) d_{2}(\theta_i,\theta_j)$. In our evaluations, we set $d_1$ to the squared Euclidean distance function, and $d_2$ the squared difference between average session lengths. In the evaluations, we test four different $\lambda_d$ weights : $0.9$, $0.99$, $0.999$ and $1$. $\lambda_d = 1$ corresponds to a placement without any non-spatial attributes \citep{Lahderanta19}. We scale the distances to the range $[0,1]$.


Each clustering is evaluated with regard to the spatial proximity between ESs (i.e., cluster centers) and APs, and the session length similarity among the ESs. We explore the means of weighted distances between servers and base stations, and the standard deviations of session lengths in each server. 

\subsection{Scenario 1: Clustering with non-spatial attribute and outlier feature}

In this section we analyze the impact of four different weights $\lambda_d$ for the spatial distance measure and the impact of outliers. In the scenario we have in total of eight different setups for the placement. 

\begin{figure}
    \centering
    \includegraphics[width = 0.99\textwidth]{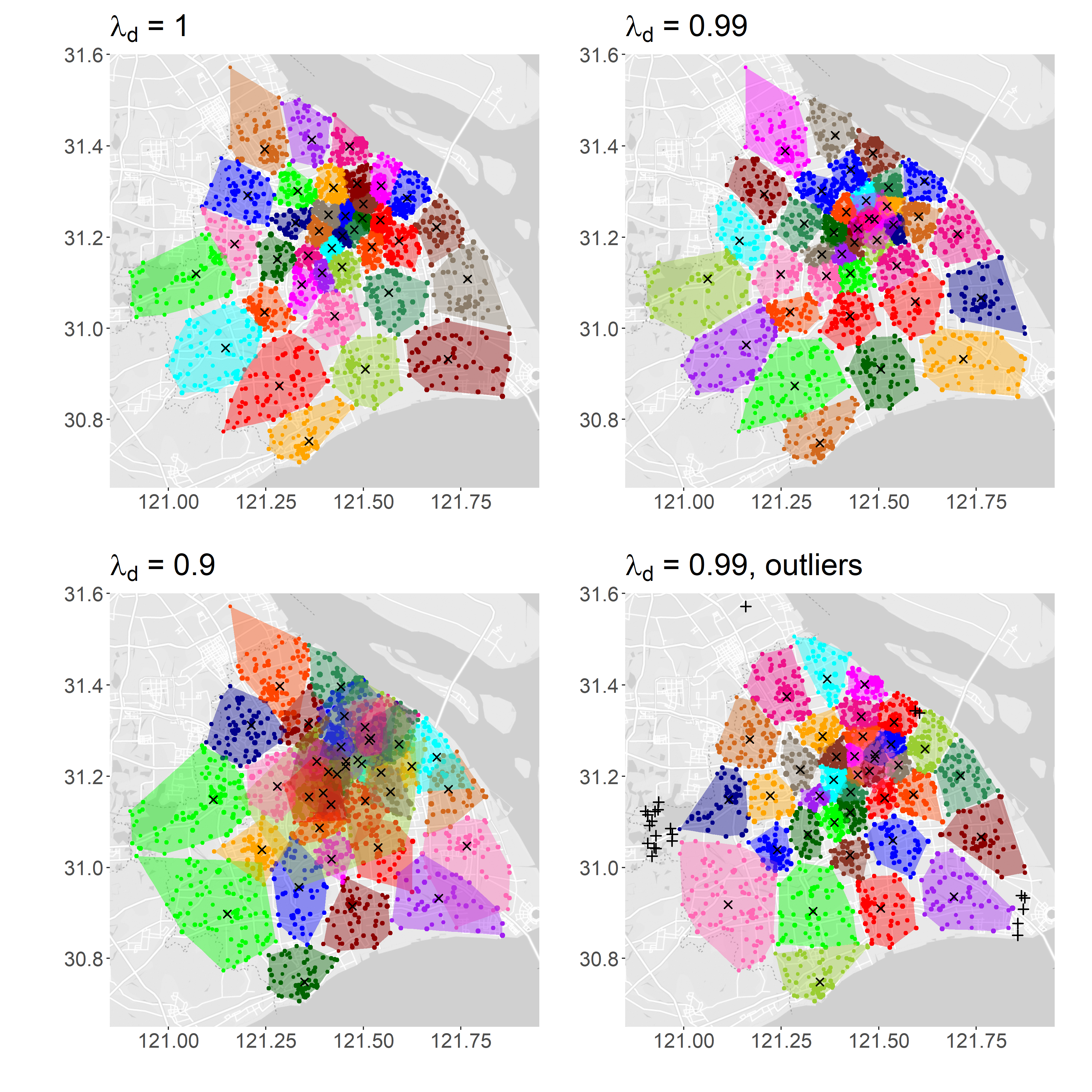}
    \caption{\textbf{The optimal placement of 38 edge servers with selected four setups.} Colors represent a set of APs that are covered by a single edge server, the edge servers are marked as crosses and outliers as plus-signs. The lower the $\lambda_d$, the more the regions overlap spatially.}
    \label{shanghai_addattr}
\end{figure}

In Figure \ref{shanghai_addattr} we have selected four setups for closer inspection. The impact of the weight value $\lambda_d$ is clearly seen as spatially overlapping server regions: As we increase the weight value of the non-spatial distance measure, that is, the value of $\lambda_d$ decreases, the resulting server regions overlap with each other more. This can be seen in the bottom left panel of Figure \ref{shanghai_addattr}, where for example the pink and purple regions on the south-east have isolated points in each others regions. 

\begin{figure}
    \centering
    \includegraphics[width = 0.8\textwidth]{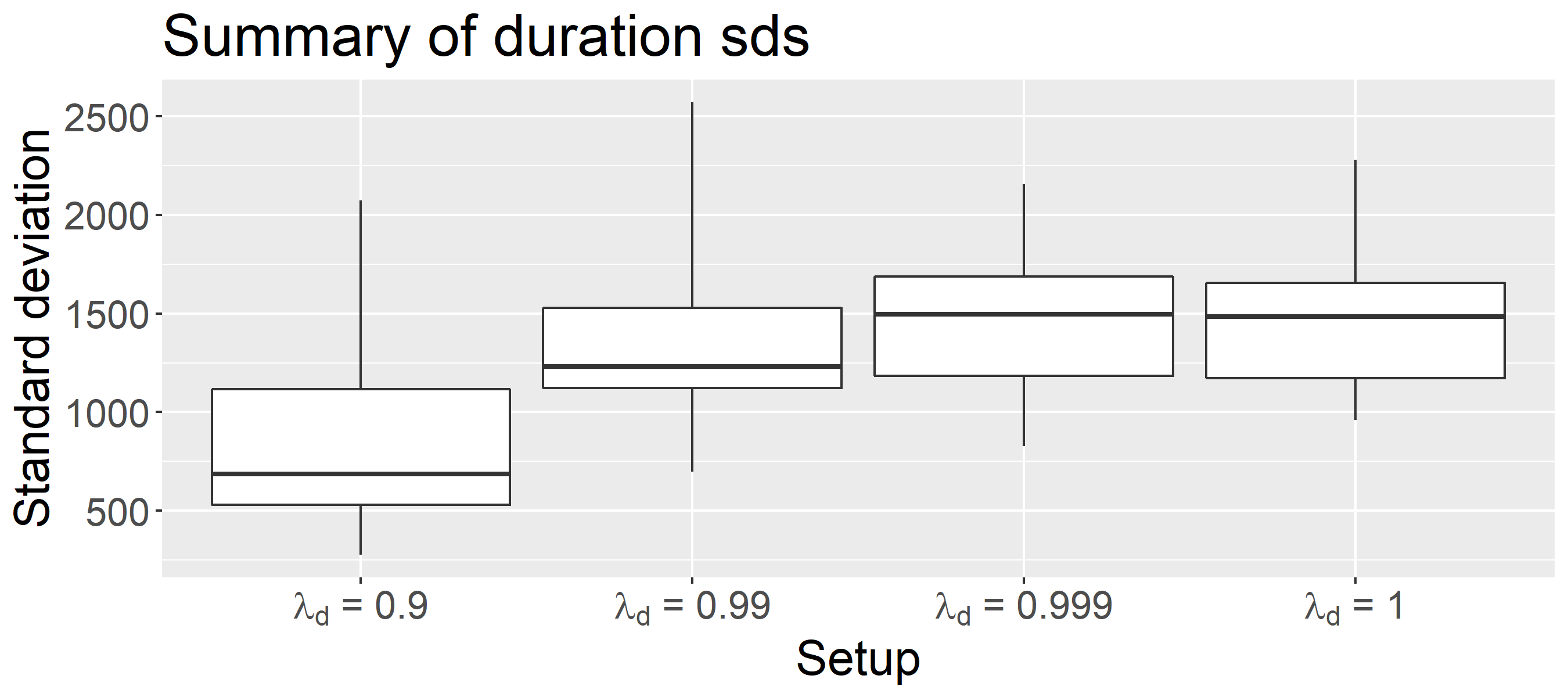}
    \caption{Distribution of standard deviations in each setup without outliers.}
    \label{shanghai_durationsds}
\end{figure}

Meanwhile, without outliers, the mean of the server-wise session duration standard deviations is cut down by 45\% when compared to the placement without non-spatial attributes (Table \ref{table_nonspat_out}). On the other hand, spatial proximity worsens by 35\%.  With outliers, the reduction in the standard deviation is as high as 53\%, while proximity degrades only 18\%. This observation is further detailed in Figure ~\ref{shanghai_durationsds}, where a trend in the standard deviations can be seen. When $\lambda_d$ is decreased, the session lengths in each cluster are more similar. 

Including outliers improves the placement in each setup when compared to not allowing outliers. Mean distances to server are slightly smaller in each setup, and the standard deviation of the session length has decreased. With $\lambda = 0.99$, both proximity and similarity are improved, by 2\% and 13\% respectively (Table~\ref{table_nonspat_out}).    

\begin{table}[ht]
\centering
\begin{tabular}{l|rr|rr}
  \toprule
  \multicolumn{1}{l|}{ } & 
  \multicolumn{2}{l|}{\textbf{Spatial proximity}} &  
  \multicolumn{2}{l}{\textbf{Session similarity}} \\

  \multicolumn{1}{l|}{ } & 
  \multicolumn{2}{l|}{(distance to ES)} &  
  \multicolumn{2}{l}{(sd. of session duration)} \\
  \midrule
\textbf{Setup} & Mean & Impr. & Mean & Impr.\\
\midrule
  $\lambda = 1$ \citep{Lahderanta19} & 3.31 & (baseline) & 1474 & (baseline)  \\ 
  \midrule
  $\lambda = 0.999$ & 3.29 & \cellcolor{green!40} 1\% & 1476 & 0\% \\ 
  $\lambda = 0.99$ & 3.36 & \cellcolor{red!20}-2\%  & 1334 & \cellcolor{green!40} 9\% \\ 
  $\lambda = 0.9$ & 4.02 & \cellcolor{red!20} -34\% & 809 & \cellcolor{green!40} 45\% \\ 
  $\lambda = 1$, outliers & 3.25 & \cellcolor{green!40}2\% & 1500 & \cellcolor{red!20} -2\%\\ 
  $\lambda = 0.999$, outliers & 3.24 & \cellcolor{green!40}2\% & 1474 & 0\% \\ 
  $\lambda = 0.99$, outliers & 3.24 & \cellcolor{green!40}2\% & 1280 & \cellcolor{green!40} 13\% \\ 
  $\lambda = 0.9$, outliers & 3.89 & \cellcolor{red!20}-18\% & 698 & \cellcolor{green!40} 53\%\\ 
\bottomrule
\end{tabular}
\caption{\textbf{Summary statistics for Scenario 1.} improvements are marked in greed, while degradations are in red.}
\label{table_nonspat_out}
\end{table}

\subsection{Scenario 2: Placement with fixed server locations}


In this scenario, we place a total of 38 edge servers in which session length is taken into account, with ten of the server locations fixed in four alternative ways, corresponding to the following setups:
\begin{enumerate}
    \item No fixed cluster center locations. This setup corresponds to the baseline placement case studied by \citet{Lahderanta19} and is identical to the setup in previous section with $\lambda = 1$.    
    \item The locations of the ten edge servers server are fixed. This corresponds to the deployment of 28 servers, with ten servers already deployed. This setup corresponds to the method studied by \citet{Loven2020a}.
    \item There's a release cost to the ten servers such that if the benefit (i.e., reduction in the objective function (\ref{lossOurs})) of relocating any of those servers is greater than the release cost, then that server is relocated. 
    \item A location preference is set for the ten fixed server location candidates. Those locations are considered more important and attract the servers more than the other locations. Location preference in edge server placement would correspond to a situation where, for example, mobile connectivity in a certain area would be above that of the surrounding areas and especially good at the center of that area. Such an area, and especially its center, would thus be a preferred location for an edge server. We set the value of the additional weight to $\gamma_i = 200$ for each preferred server location $i=1,\dots,10$. This value is larger than the largest weight value (i.e. 156) in the data set.
\end{enumerate}

\begin{figure}
    \centering
    \includegraphics[width = 0.99\textwidth]{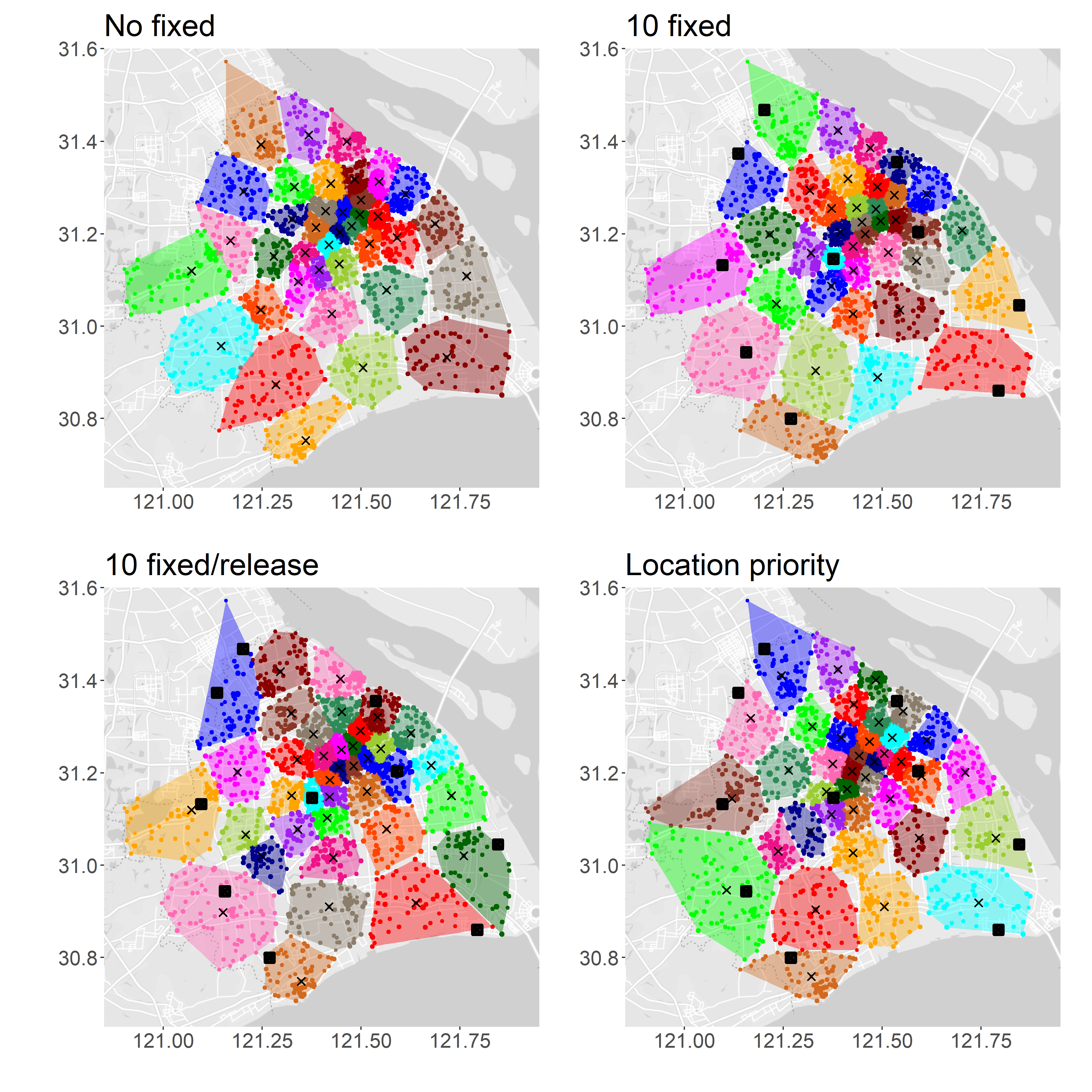}
    \caption{\textbf{The optimal placement of 38 edge servers with selected four setups.} Colors represent a set of APs that are covered by a single edge server, the edge servers are marked as crosses and the fixed servers as squares.}
    \label{shanghai_predet_centers}
\end{figure}

\begin{table}[ht]
\centering
\small
\begin{tabular}{l|rr|rr}
  \toprule
  \multicolumn{1}{l|}{ } & 
  \multicolumn{2}{l|}{\textbf{Spatial proximity}} &  
  \multicolumn{2}{l}{\textbf{Session similarity}} \\
  \multicolumn{1}{l|}{ } & 
  \multicolumn{2}{l|}{(distance to ES)} &  
  \multicolumn{2}{l}{(sd. of session length)} \\
  \midrule
\textbf{Setup} & Mean & Impr. & Mean & Impr. \\
  \midrule
  \midrule
  Fixed \citep{Loven2020a}& 3.70 & (baseline) & 1480 & (baseline)\\ 
  \midrule
  Fixed, $\lambda_d$ & 3.79 & \cellcolor{red!20}-2\% & 1228 & \cellcolor{green!40} 17\% \\ 
  Fixed/release & 3.46 & \cellcolor{green!40}6\% & 1485 & 0\%  \\ 
  Fixed/release, $\lambda_d$ & 3.56 & \cellcolor{green!40}4\%  & 1424 & \cellcolor{green!40}4\%  \\ 
  \midrule
  \midrule
  No fixed \citep{Lahderanta19} & 3.31 & (baseline) & 1474 & (baseline) \\ 
  \midrule
  Loc. preference & 3.28 & \cellcolor{green!40}1\% & 1483 & \cellcolor{red!20}-1\% \\ 
  Loc. preferenc, $\lambda_d$ & 3.36 & \cellcolor{red!20}-2\% & 1273 & \cellcolor{green!40}14\% \\ 
   \bottomrule
\end{tabular}
\caption{\textbf{Summary statistics for Scenario 2.} Improvements are marked in green, while degradation is in red.}
    \label{shanghai_table}
\end{table}

%


Furthermore, we incorporate the session length attribute to the placement. In this scenario a weight value for spatial attributes $\lambda_d = 0.99$ is used with each setup mentioned above, so that in total of eight setups are evaluated.

Fig. \ref{shanghai_predet_centers} compares these four different setups without non-spatial attributes. Overall, each setup produces deployment of servers with some important differences. First, the benefit of relocating the fixed servers is clear: some of the fixed servers in the south-west and in the south-east are located on the edge of the map which overall worsens the average distance to the center, that is, latency to the server.

Table \ref{shanghai_table} further details the changes in proximity and similarity. Compared with a baseline method \citep{Loven2020a}, fixed servers with the non-spatial attributes indeed improves similarity by 17\% while slightly (-2\%) degrading in proximity. Just allowing the releasing of fixed servers, on the other hand, only improves on proximity (+6\%). However, jointly considering both the non-spatial attributes as well the releasing of fixed servers improves on both proximity and similarity by 4\%. On the other hand, the center-based location preference, compared to a baseline method \citep{Lahderanta19}, affects proximity but little. Including the non-spatial attributes, however, improves on similarity by 14\% while reducing proximity just slightly (-2\%). 

Overall, average latency is improved as we allow the servers to be relocated. The setup with location preference can be seen as a ``relaxed'' version of the setup with release cost on pre-determined center locations, as it pulls the servers closer to the predetermined locations rather than forcing the servers to be located exactly at the predetermined locations. Furthermore, the mean and median distances in setups with location preference are much smaller than in the other setups. When it comes to setups with average session length, the non-spatial attribute improved session similarity as expected, while still maintaining reasonable spatial proximity in each setup.

All the four setups produce clusterings with no unnatural cluster structure. The differences in the resulting clusters are mainly due to the locations of the predetermined spots: forcing the servers to the fixed locations can cause worse average latency in the deployment, which can be solved by allowing servers at the fixed locations to relocate.


\section{Discussion and Conclusion}
\label{discussion}

In this article, we introduced new extensions to the PACK method for capacitated clustering. The new extensions included support for multidimensional attributes for points of interest, outliers, releasing fixed centers, and setting a location preference for cluster centers. PACK method supports the joint consideration of all the extensions.




Evaluating PACK with edge server placement, we demonstrated the use of non-spatial attributes together with outliers and pre-determined server locations. 
We used a non-spatial attribute, the average session duration, as a proxy to AP application usage profiles. As result, the average standard deviation of the edge server session durations was reduced at best 53\%. Further, the inclusion of outliers impacted the placement in a positive way, decreasing both the mean distances between a server and a base station and the standard deviations of session length in each server. Compared against the location preference extension which favored the neighborhoods of the given points, the proposed releasing of fixed centers emphasized only the given points, either locating a center exactly on them or allowing a free placement for the center.

In conclusion, we proposed PACK method for capacitated spatial clustering, with number of extensions on cluster center locations, outliers and non-spatial attributes. We demonstrated these extensions in an edge server placement problem, where two different scenarios were examined. We concluded that the inclusion of non-spatial attributes allows a greater control between spatial proximity and attribute similarity. Further, a better overall proximity and similarity was achieved when outlier points were detected and discarded from the cluster analysis. Control over the cluster centers provided a way to apply preference to certain locations without affecting the overall cluster quality. We argue that the inclusion of the novel extension and the flexibility of the PACK algorithm, combined with the easy-to-use open-source R software package, provides a versatile toolbox spatial clustering, especially in relation to edge server placement.


\noindent\textbf{Funding.}
This research is supported by Academy of Finland 6Genesis Flagship (grant 318927), the Infotech Oulu research institute, the Future Makers program of the Jane and Aatos Erkko Foundation and the Technology Industries of Finland Centennial Foundation, by Academy of Finland PROFI5 funding for mathematics and AI: data insight for high-dimensional dynamics, and by the personal grant for Lauri Lovén on Edge-native AI research by the Tauno Tönning foundation.

\noindent\textbf{Conflicts of interest.}
The authors declare that they have no conflict of interest.

\noindent\textbf{Availability of data and material.}
The data set of the experiments is available at http://sguangwang.com/TelecomDataset.html.

\noindent\textbf{Code availability.}
\texttt{rpack} source code is freely available at \\ https://github.com/terolahderanta/rpack.

\bibliographystyle{spbasic}
\bibliography{refs}







\end{document}